\documentclass[journal]{./IEEEtran}

\usepackage{cite}
\ifCLASSINFOpdf
  \usepackage[pdftex]{graphicx}
  \DeclareGraphicsExtensions{.pdf,.jpeg,.png}
\else
  \usepackage[dvips]{graphicx}
  \DeclareGraphicsExtensions{.eps}
\fi

\usepackage[cmex10]{amsmath}
\usepackage{amsfonts,amssymb,amsthm,mdwmath}

\usepackage{array}

\ifCLASSOPTIONcompsoc
  \usepackage[caption=false,font=normalsize,labelfont=sf,textfont=sf]{subfig}
\else
  \usepackage[caption=false,font=footnotesize]{subfig}
\fi
\usepackage{stfloats}
\usepackage{url}
\usepackage{bbm}
\usepackage{color}
\usepackage{acronym}
\usepackage{algorithm}
\usepackage{algorithmic}

\begin{document}
\acrodef{PPP}[PPP]{Poisson point process}
\acrodef{MPLP}[MPLP]{Manhattan Poisson line process}
\acrodef{MPLCP}[MPLCP]{Manhattan Poisson line Cox process}
\acrodef{CDF}[CDF]{cumulative distribution function}
\acrodef{PDF}[PDF]{probability distribution function}
\acrodef{RV}[RV]{random variable} 
\acrodef{i.i.d.}[i.i.d.]{independent and identically distributed}
\acrodef{MGF}[MGF]{moment generating function}
\acrodef{kNN}[kNN]{$k$ nearest-neighbors}
\acrodef{POI}[POI]{points of interest}
\acrodef{VANET}[VANET]{vehicular ad hoc network}
\acrodef{SNR}[SNR]{signal-to-noise ratio}
\acrodef{1D}[1D]{one-dimensional}
\acrodef{RSU}[RSU]{road side unit}
\acrodef{NLoS}[NLoS]{non-light-of-sight}

\title{On the $k$ Nearest-Neighbor Path Distance from the Typical Intersection in the Manhattan Poisson Line Cox Process}
\author{Konstantinos Koufos, Harpreet S. Dhillon, Mehrdad Dianati and Carl P. Dettmann}  

\maketitle

\begin{abstract}
In this paper, we consider a Cox point process driven by the Manhattan Poisson line process. We calculate the exact cumulative distribution function (CDF) of the path distance (L1 norm) between a randomly selected intersection and the $k$-th nearest node of the Cox process. The CDF is expressed as a sum over the integer partition function $p\!\left(k\right)$, which allows us to numerically evaluate the CDF in a simple manner for practical values of $k$. The distance distributions can be used to study the $k$-coverage of broadcast signals transmitted from a \ac{RSU} at an intersection in intelligent transport systems (ITS). They can also be insightful for network dimensioning in urban vehicle-to-everything (V2X) systems, because they can yield the exact distribution of network load within a cell, provided that the \ac{RSU} is located at an intersection. Finally, they can find useful applications in other branches of science like spatial databases, emergency response planning, and districting. We   corroborate the applicability of our distance distribution model using the map of an urban area. 
\end{abstract}


\section{Introduction}
The development of the road network is a key component of urban planning because it greatly affects commuting efficiency, districting, emergency response dispatching, and first-aid services, to name but a few. Since the recent advent of wireless connectivity for pedestrians, and the ongoing proliferation of connected vehicles through vehicle-to-everything (V2X) systems, the road network is also the setting for several location-based e-services~\cite{Karagiannis2011}. Exemplar applications could be electric vehicles querying over the internet for the nearest charging stations, and/or pedestrians searching with their smartphones for the closest available taxis~\cite{Papadias2003, Jensen2003, Mouratidis2006, Hua2018}. 

\subsection{Modeling road networks}
The simplest models for urban road networks utilize just a set of vertices and edges~\cite{Thomson1995, Marshall2018}. The vertices may represent junctions, the start/end points of roadways, critical locations where the speed limit or the travel direction changes, etc. Naturally, two vertices are connected by an edge if there is a straight link between them, giving rise to the adjacency matrix of the road network, which is not necessarily binary. The graph representation is enhanced by assigning weights to the edges, which might be proportional to the (average or minimum) travel time and fuel cost, along the road segment(s) that the edge represents. Algorithms exploring the graph have been also developed, e.g., the best-first search to identify the nearest neighbors from a vertex and Dijkstra's algorithm to find the shortest paths, i.e., the sequence of edges of minimal aggregate cost between two non-adjacent vertices~\cite{Knuth1998}. We will also utilize Dijkstra's algorithm with edge weights equal to the Euclidean distance of road segments while validating our distance distribution models with a real map. 

Another line of research, particularly useful for emergency response planning, assumes that the edge weights are proportional to the length of the associated streets, and models random events along the edges. If these events represent points of emergency, the graph distance distributions, see~\cite{Wei2019}, can reveal the intrinsic properties of the response system we need to build to combat all emergencies effectively. For instance, they can be used to infer the number of ambulances, medical personnel, etc. we have to deploy.

While certainly important, the graph-based approaches apply to specific road networks. Even though different cities can share common road graph properties~\cite{Barthelemy2011}, the graph-based models provide limited abstraction. Besides, due to the high complexity of graph-based routines, it is often impossible to model the road network very precisely. Also, the graph representation cannot answer questions pertinent to network planning, e.g., what is the minimum required intensity of charging stations, so that two of them are within a driving distance of one kilometer from a randomly chosen intersection, with probability at least $90 \%$?  This paper aims to bridge this gap. We argue that the mathematical tools of stochastic geometry, see~\cite{Stoyan2013} for an introduction, widely and successfully utilized during the past 15 years in the performance evaluation of random wireless communication networks, see~\cite{Andrews2011, Baccelli2018a, Baccelli2018b, Chetlur2018, Chetlur2018b,Koufos2019b,Koufos2018b}, can also be insightful for urban road planning.

Unlike the graph-based methodology, the stochastic geometry framework does not consider a specific road network. Only the intensity of streets is available (or can be estimated). In addition, we assume that: (i) The road layout has a relatively regular structure, hence, a \ac{MPLP} is a realistic model for it, and (ii) the locations of \ac{POI} or facilities, e.g., gas stations, events triggering police action, etc. follow a homogeneous \ac{PPP} along each street. Under these assumptions, i.e., a \ac{MPLCP} for the locations of \ac{POI}, we will derive the path distance (L1 norm) distribution of the \ac{kNN} (or \ac{POI}) from a randomly selected intersection. We have identified three potential applications, namely spatial database queries, districting, and urban \acp{VANET}, where the \ac{kNN} path distance distributions could be of use. We elaborate on them next.

\subsection{Motivation and prior art}
In the \ac{kNN} query, a spatial database returns the locations of the $k$ nearest objects, in terms of network distance, to the query point (or agent)~\cite{Papadias2003}. Consider, for instance, a driver looking for the $k$ nearest hotels (static objects) in terms of travel time, or a pedestrian querying for the $k$ nearest vacant cabs (mobile objects). The agent reports its location, and the database solves the query using, e.g., a graph representation for the road network~\cite[Fig.~2]{Jensen2003}. 

Even with static objects, the graph dynamically changes due to varying traffic conditions, and the computational complexity can quickly explode, especially with frequent queries from mobile agents~\cite{Mouratidis2006}. The server must continuously track and update the locations of the $k$ nearest objects for all the agents. Because of that, neglecting the constraints imposed by the road network, and using just Minkowski distances to solve the \ac{kNN} problem, especially for group queries, has not been abandoned~\cite{Hua2018}. The study in~\cite{Papadias2003} has pointed out that the Euclidean distance is a lower bound to the network distance, and thus, we could use it to prune the search space in \ac{kNN} queries. However, pruning based on a lower bound is not always effective. Therefore the calculation of the exact path distance distributions, as we will do in this paper, will be very helpful. 

Apart from spatial databases, the \ac{kNN} distance distributions can also be used in the planning of dispatching policies for emergency response services and districting~\cite{Mercado2020}. In balanced district design, the road network of a metropolitan area is partitioned into smaller units (territories) which contain about the same expected number of road accidents so that the workload is divided equally among police departments~\cite{Mercado2020}. Given the size of the districts, the \ac{kNN} distributions can be used to calculate the probability that a police department can cover the $k$ nearest emergencies under some  probabilistic target time guarantees. 

The ongoing deployment of 5G networks and the standardization activities for V2X communication, e.g., the transmission of cooperative awareness messages from the vehicles to the infrastructure~\cite{ETSICAM}, and the response of the latter with the collective perception message~\cite{ETSICPM} have motivated the development of spatial models tailored to vehicular networks. Despite the fact that urban streets have finite lengths, line processes have been extensively used in wireless communication research to gain analytical insights into the network's performance~\cite{Baccelli2018a, Baccelli2018b, Chetlur2018, Chetlur2018b}. The one- and two-dimensional \acp{PPP} are valid models for urban \acp{VANET}, in the high- and the low-reliability regime respectively~\cite{Haenggi2017}. As a result, the optimal transmission probability in \acp{VANET} modeled by Cox point processes is different than that calculated using the \ac{PPP}~\cite{Chetlur2018b}. To draw valid conclusions about the network's performance the road intersections must be modeled explicitly. For motorway \acp{VANET}, on the other hand, the superposition of \ac{1D} point processes is sufficient~\cite{Koufos2019, Koufos2020, Koufos2018}.  

The studies in~\cite{Baccelli2018a,Baccelli2018b,Chetlur2018} have used a Poisson line process to capture the random orientation of streets, and stationary \ac{1D} \acp{PPP} to model the locations of vehicles per street. In the resulting Poisson line Cox process $\Phi_c$, the study in~\cite{Baccelli2018a}  has evaluated the distance distribution between an arbitrary point in the plane and the nearest point of $\Phi_c$. This is essentially the serving distance distribution in a cellular  vehicular network with nearest base station association, where the locations of base stations follow the two-dimensional \ac{PPP}~\cite{Baccelli2018b}. The study in~\cite{Chetlur2018} has derived the coverage probability for the typical receiver in a  \ac{VANET} and pointed out the conflicting effect of the intensities of the roads and vehicles. It has also solved for the distance distribution between the typical vehicle and the nearest vehicle of $\Phi_c$. The study in~\cite{Baccelli2015} has used the \ac{MPLCP} to model the locations of base stations in urban street microcells and calculated their distance distribution to the origin to study properties of the interference distribution. Finally, for some recent results on the reliability of inter-vehicle communication in urban streets of finite length, forming intersections and T-junctions, see~\cite{Haenggi2020}.

Unfortunately, the above studies have measured the distances in the Euclidean (L2 norm) sense, even though the attenuation of wireless signals, especially in millimeter-wave frequencies, is better described by a street canyon model~\cite[Eq.~(1)]{Heath2018}. In this regard, the \ac{kNN} Manhattan distance distributions will be useful in investigating the $k$-coverage of wireless signals, diffracted around buildings at road intersections as they propagate. We will use them to  identify, e.g., how many vehicles within half a kilometer from an intersection can successfully receive broadcast safety messages with probability at least $q \%$? Thus far, the \ac{kNN} distributions have been identified for Poisson and binomial processes see~\cite{Evans2002,Moltchanov2012,Afshang2017}, without considering the deployment constraints due to the road layout. For $k\!=\!1$, more general convex geometries like the $n$-sided polygon have been also investigated~\cite{Durrani2013}.

\subsection{Contributions}
The \ac{PDF} of the shortest path between a random intersection and a point of the \ac{MPLCP} has been recently calculated in~\cite{Chetlur2020}. In this paper, we will generalize this result to $k\!\geq\!1$ nearest points. We present various methods to do that, and finally, we cast the solution as a sum over the integer partitions of $k$. The computational complexity of the suggested numerical algorithm is low, and the \acp{CDF} can be easily obtained for practical  values of $k$. Finally, it should be noted that the \ac{CDF} of the distance between a random intersection and the $k$-th nearest point of the \ac{MPLCP} can serve as a lower bound to the \ac{CDF} of the distance between a random position of the road and the $k$-th nearest neighbor of the \ac{MPLCP}. For $k\!=\!1$ both \acp{CDF} are computed in~\cite{Chetlur2020}. 

Section~\ref{sec:System} formally introduces the \ac{MPLCP}. Section~\ref{sec:Lt} calculates the \ac{PDF} of the total length $L_t$ of line segments inside a Manhattan square, centered at a randomly selected road intersection. In Section~\ref{sec:MGF}, we calculate the \ac{MGF} of the \ac{RV} $L_t$, and in Section~\ref{sec:Integer} we present a numerical algorithm which can be used to compute the \ac{CDF} of the distance between an intersection and the $k$-th nearest point of the \ac{MPLCP}. In Section~\ref{sec:Numericals}, we validate the suggested algorithm against simulations. We also use real data for the road network obtained from an urban area to demonstrate its improved performance against existing models based on the PPP. In Section~\ref{sec:Conclusions}, we conclude.

\section{System model and notation}
\label{sec:System}
A line process, in layman's terms, is just a random collection of lines. If we limit our attention to undirected lines in the Euclidean plane, each line $\ell_i$ can be uniquely determined by the following parameters: the length $\rho_i\!\geq\!0$ and the angle $\phi_i\!\in\!\left[-\pi,\pi\right]$, measured counter-clockwise, of the line segment being perpendicular to line $\ell_i$ and passing though the origin~\cite[Chapter~8.2.2]{Stoyan2013}. Therefore a line process can be associated with a point process, and vice versa, where the line $\ell_i$ is uniquely mapped to the point $x_i\!\in\!\mathbb{R}^2$ with polar coordinates $\left(\rho_i,\phi_i\right)$. The associated point process is often called the representation space of the line process.  

Let us consider the realizations of two independent \ac{1D} \acp{PPP} of equal intensity $\lambda$, along the $x$ and $y$ axes, and construct the associated realization of vertical and horizontal lines. All points on the $x$ axis give rise to vertical lines $\phi_i\!\in\! \left\{0,\pi\right\}$, and all points along the $y$ axis correspond to horizontal lines $\phi_i\!\in\! \left\{-\pi/2,\pi/2\right\}$. This is known as the Manhattan Poisson line process (MPLP). It is a stationary and motion-invariant line process owing to the stationarity and motion-invariance of the \ac{PPP} in the representation space. Its intensity, defined as the mean total length of lines per unit area, is equal to $2\lambda$~\cite[Chapter~8.1]{Stoyan2013}. 

Due to the fact that the contact distribution of the \ac{1D} \ac{PPP} is exponential, the distances between neighboring intersections of a \ac{MPLP} follow the exponential distribution too with rate $\lambda$. The set of horizontal lines is denoted by $\Phi_{lh}\!=\!\left\{L_{h_1}, L_{h_2},\ldots\right\}$, the set of vertical lines by $\Phi_{lv}\!=\!\left\{L_{v_1}, L_{v_2},\ldots\right\}$, and $\Phi_l\!=\!\left\{\Phi_{lh},\Phi_{lv}\right\}$ is the resulting \ac{MPLP}. 

Let us assume that along each line there are facilities, e.g. gas stations, whose locations follow another \ac{1D} \ac{PPP} of intensity $\lambda_g$. Conditionally on the realization of the line process, the locations of facilities are independent. Under these assumptions, the distribution of facilities becomes a stationary Cox point process, denoted by $\Phi_g$, and driven by $\Phi_l$. A Cox point process is in general a doubly-stochastic \ac{PPP} where the intensity measure is itself random, and it is subsequently constructed in a two-step random mechanism. See~\cite[Chapters 3 and 4]{Dhillon2020} for an introduction. In our case, a set of random lines parallel to the $x,y$ axes is generated first, followed by the random locations of facilities along each line. The intensity of the \ac{MPLCP} is $2\lambda\lambda_g$~\cite{Stoyan2013}. 

In this paper, we will calculate the path distance (L1 norm or Manhattan distance) distribution between an arbitrary intersection of the \ac{MPLP} and its $k$-th nearest facility. In our calculations, the width of the road is ignored, because it is negligible as compared to the expected distance between neighboring road intersections $\lambda_g^{-1}$. All roads are assumed bi-directional, and the locations of the facilities are constrained along the road network. Note that it is straightforward to extend the calculations for different intensities $\lambda_h, \lambda_v$ between the vertical and horizontal streets. Consider for instance few main vertical streets traversing the city and many horizontal side streets. Unless otherise stated, we will use $\lambda_h\!=\!\lambda_v\!=\!\lambda$ for presentation clarity.   

\begin{figure}[!t]
 \centering
\includegraphics[width=3in]{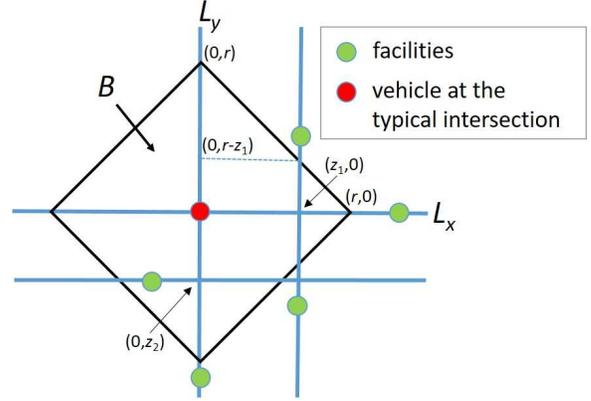}
 \caption{Example realization of a \ac{MPLCP} of facilities.}
 \label{fig:Grids}
\end{figure}

Owing to the stationarity of the \ac{MPLP}, we can add an intersection at the origin of the $x-y$ plane, and two (undirected) lines $L_x, L_y$ passing through it, and aligned with the $x$ and $y$ axis respectively. Therefore under Palm probability, the resulting line process becomes $\Phi_L\!=\! \left\{\Phi_l\cup\left\{L_x,L_y\right\}\right\}$, and the point process of facilities is the superposition of the point process $\Phi_g$ and the two \acp{PPP} of intensity $\lambda_g$ along $L_x$ and $L_y$. See Fig.~\ref{fig:Grids} for an illustration. Due to Slivnyak's Theorem~\cite[Chapter~8.2]{Stoyan2013}, the lines $L_x, L_y$ does not affect the distribution of $\Phi_l$. As a result, the distance distribution between a randomly selected intersection of $\Phi_l$ and its $k$-th nearest facility, is essentially equal to the distribution between the origin of the augmented grid $\Phi_L$ and its $k$-th nearest facility. 

The \ac{CDF} for the path distance of the $k$-th nearest facility to the origin is denoted by $F_{R_k}\!\!\left(r\right)\!=\!\mathbb{P}\!\left(R_k\!\leq\!r\right)\!=\! 1-\mathbb{P}\!\left(R_k\!>\!r\right)$. The complementary \ac{CDF}, $\mathbb{P}\!\left(R_k\!>\!r\right)$, is equal to the sum of the probabilities $P_j, j\in\left\{0,1,2, \ldots \left(k\!-\!1\right)\right\}$, hence, $F_{R_k}\!\left(r\right) = 1 - \sum\nolimits_{j=0}^{k-1} P_j$, where $P_j$ is the probability that there are exactly $j$ facilities inside the square, see Fig.~\ref{fig:Grids}, which is the locus of points with Manhattan distance $r$ to the origin. The set of all points inside the square, including the sides, is denoted by $B\!\left(r\right)\!\equiv\! B$ for brevity.

\noindent
The \ac{CDF} of the \ac{RV} $R_1$ has been derived in~\cite[Theorem 1]{Chetlur2020}  
\begin{equation}
\label{eq:R1}
F_{R_1}\!\left(r\right) = 1 - P_0 = 1 - e^{-4 \lambda_g r} \, e^{-4\lambda r\left(1-a_0\right)}, 
\end{equation}
where $a_0=\frac{1-e^{-2\lambda_g r}}{2\lambda_g r}$ and $P_0\!\triangleq\! e^{-4 \lambda_g r} \, e^{-4\lambda r\left(1-a_0\right)}$ is the probability that no facility lies in $B$. 

In addition, we define the \ac{RV} $N_p\!\left(\Phi\cap B\right)$ which counts the number of points of the Cox process, driven by the line process $\Phi$, within $B$. The total number of lines intersecting $B$, excluding the typical lines $L_x, L_y$ is denoted by the \ac{RV} $N$. Furthermore, the \ac{RV} $L_i\!\geq\! 0$ describes the random length of the $i$-th line $\ell_i$ intersecting $B$, and the \ac{RV} $L_t\!\geq\!4r$ describes the total length of line segments in $B$ including the contribution, $4r$, due to the typical lines. Finally, the realizations of the \acp{RV} $L_i$ and $L_t$ are both denoted by $l$.

\section{Calculating $P_k$ using the distribution of $L_t$}
\label{sec:Lt}
Given the realization $l$ of the \ac{RV} $L_t$, the number of facilities in $B$ follows a Poisson distribution with parameter $\lambda_g l$, ${\text{Po}}\!\left(\lambda_g l\right)$. As a result, based on the law of total expectation, the probability $P_k$ that there are $k$ facilities in $B$ can be obtained by averaging the Poisson distribution over the \ac{PDF}, $f_{L_t}\!\left(l\right)$, of the total length of line segments $L_t$ in $B$. Hence, 
\begin{equation}
\label{eq:Pkg}
P_k = \int\nolimits_{4r}^\infty \frac{e^{-\lambda_g l}\left(\lambda_g l\right)^k}{k!} f_{L_t}\!\left(l\right) {\rm d}l, 
\end{equation} 
where the lower integration limit equals $4r$, because $B$ always contains the segments due to the typical lines $L_x, L_y$. 

In order to derive the \ac{PDF} $f_{L_t}\!\left(l\right)$, we start with the random number $N$ of line segments intersecting $B$, which follows the Poisson distribution with parameter $4\lambda r$. Recall that $\lambda$ is the density of intersection points along a line, and $2r$ is the length of the diagonal of $B$. Conditionally on $N\!\geq\!1$, the abscissas (ordinates) of the line segments parallel to $L_y$ ($L_x$) are distributed uniformly at random in $\left(-r,r\right)$. As a result, the distribution of the \ac{RV} $L_i$ describing the length of the $i$-th line segment in $B$ is uniform too. In order to derive its CDF, we note that $L_i$ takes values in $\left(0,2r\right)$ and thus, $\mathbb{P}\left(L_i\leq l\right) \!=\! \frac{l}{2r}, \, l\!\in\!\left(0,2r\right)$. For instance, the vertical line with abscissa $z_1$ in Fig.~\ref{fig:Grids} has length $l\!=\!2\left(r-z_1\right)$ in $B$. 

Conditionally on the realization $n\geq 1$ for the \ac{RV} $N$, the total length of line segments in $B$, $\sum\nolimits_{i=1}^n L_i$, is equal to the sum of $n$ \ac{i.i.d.} uniform \acp{RV} in $\left(0,2r\right)$. As a result, the sum $\sum\nolimits_{i=1}^n L_i$ follows the Irwin-Hall distribution with PDF 
\begin{equation}
\label{eq:Ln}
\sum\limits_{i=1}^n L_i \!\sim\! \frac{1}{2r \left(n-1\right)!} \sum\limits_{k=0}^{\left \lfloor{\frac{l}{2r}}\right \rfloor } \left(-1\right)^k \binom{n}{k} \left(\frac{l}{2r}-k\right)^{n-1}, 
\end{equation}
where $n\!\geq\! 1$ and $l\!\geq\! 0$.

In order to compute the PDF of $L_t$, we need to average equation~\eqref{eq:Ln} over the Poisson distributed number $N$ for $n\!\geq\!1$. The case $N\!=\!0$, i.e., no intersections along $\left\{L_x\cup L_y\right\}\cap B$, which occurs with probability $e^{-4\lambda r}$, leads to $L_t\!=\!4r$ and it is treated separately below. 
\begin{equation}
\label{eq:Lt}
\begin{array}{ccl}
f_{L_t}\!\left(l\right) \!\!\!&=&\!\!\!  e^{-4\lambda r} \delta_{l,4r} \, + \\ \!\!\!\!\!& &\!\!\!\!\! \!\!\!\!\!\!\!\!\!\!\mathbb{E}_{n\geq 1}\!\left\{\! \frac{1}{2r \left(n-1\right)!} \!\! \sum\limits_{k=0}^{\left \lfloor{\frac{l-4r}{2r}}\right \rfloor } \!\! \left(-1\right)^k \binom{n}{k} \left(\frac{l-4r}{2r}\!-\!k\right)^{n-1} \! \right\} \\ 
\!\!\!&=&\!\!\! e^{-4\lambda r} \delta_{l,4r} + \sum\limits_{n=1}^\infty \frac{\left(4\lambda r\right)^n e^{-4\lambda r}}{n!} \frac{1}{2r \left(n-1\right)!} \times \\ & & \sum\limits_{k=0}^{\left \lfloor{\frac{l-4r}{2r}}\right \rfloor } \left(-1\right)^k \binom{n}{k} \left(\frac{l-4r}{2r}-k\right)^{n-1},\, l\geq 4r,
\end{array}
\end{equation}
where $\delta_{x,y}\!=\!1$ for $x\!=\!y$ and $\delta_{x,y}\!=\!0$ otherwise, is the Kronecker delta function, and also note that  equation~\eqref{eq:Ln} has been shifted to the right by $4r$. 

\begin{figure}[!t]
 \centering
\subfloat[$\lambda\!=\!0.5, \, r\!=\!4$] {\includegraphics[width=3in]{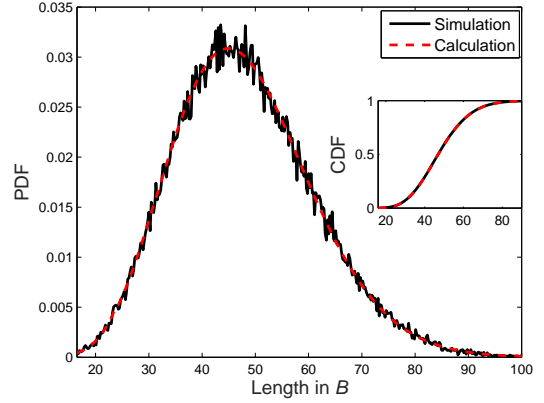}\label{fig:fLt1}} \hfil 
\subfloat[$\lambda\!=\!0.5, \, r\!=\!2$] {\includegraphics[width=3in]{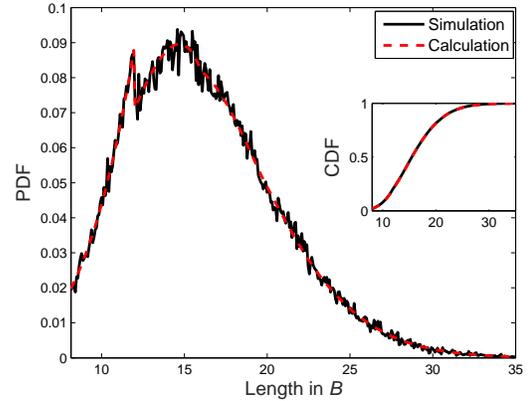}\label{fig:fLt2}}
 \caption{The \ac{PDF} and the \ac{CDF} (inset) of the total length $L_t$ of line segments in $B$. In $(b)$, the probability, $e^{-4\lambda r}$, that the only line segments are due to $L_x, L_y$ is around $2\%$, evident in the left tail of the \ac{PDF}. $10^5$ simulations.}
 \label{fig:fLt}
\end{figure}

The above expression can be simplified, to some extent, by interchanging the order of summations, and while doing so, carefully setting the lower limit of the sum with respect to $n$. $f_{L_t}\!\left(l\right)$
\begin{equation}
\label{eq:Lt2}
\begin{array}{ccl}
{} \!\!\!\!\!&=&\!\!\!\!\! e^{-4\lambda r} \delta_{l,4r} + \frac{e^{-4\lambda r}}{2r} \sum\limits_{k=0}^{\left \lfloor{\frac{l-4r}{2r}}\right \rfloor } \left(-1\right)^k \times \\ & & \sum\limits_{n=\max\left\{1,\, k\right\}}^\infty \frac{\left(4\lambda r\right)^n}{n!} \frac{1}{\left(n-1\right)!} \binom{n}{k} \left(\frac{l-4r}{2r}-k\right)^{n-1} \\ 
\!\!\!\!\!&=&\!\!\!\!\! e^{-4\lambda r} \delta_{l,4r} + \frac{e^{-4\lambda r}}{2r} \Big( 4\lambda r \,  {}_0F_1\!\left(2,2\lambda\left(l-4r\right)\right) + \\ & & \sum\limits_{k=1}^{\left \lfloor{\frac{l-4r}{2r}}\right \rfloor } \left(-1\right)^k \sum\limits_{n=k}^\infty \frac{\left(4\lambda r\right)^n}{n!} \frac{1}{\left(n-1\right)!} \binom{n}{k} \left(\frac{l-4r}{2r}-k\right)^{n-1} \Big) \\ 
\!\!\!\!\!&=&\!\!\!\!\! e^{-4\lambda r} \delta_{l,4r} + \frac{e^{-4\lambda r}}{2r} \Big( 4\lambda r\,  {}_0F_1\!\left(2,2\lambda\left(l-4r\right)\right) + \\ & & \sum\limits_{k=1}^{\left \lfloor{\frac{l-4r}{2r}}\right \rfloor } \!\! \frac{\left(-4\lambda r\right)^k}{k!} \left(\frac{l-4r}{2r}-k\right)^{k-1} {}_0F_1\left(k,4\lambda r \left(\frac{l-4r}{2r}-k\right) \right) \Big),
\end{array}
\end{equation}
where $ l\geq 4r$ and ${}_0F_1\!\left(\alpha,z\right)\!=\! \sum\nolimits_{k=0}^\infty \frac{1}{\Gamma\!\left(\alpha+k\right)}\frac{z^k}{k!}$ is the regularized hypergeometric function; example validations of equation~\eqref{eq:Lt2} are depicted in Fig.~\ref{fig:fLt}. 

The final expression in equation~\eqref{eq:Lt2} is quite complicated to use in the integral in~\eqref{eq:Pkg}, hence, calling for another approach to evaluate the probabilities $P_k$. 

\section{Calculating $P_k$ using MGFs}
\label{sec:MGF}
Since the \ac{PDF} of the \ac{RV} $L_t$ has a complicated form, we may instead work with its \ac{MGF}, $M_{L_t}\!\left(t\right) = \mathbb{E}\!\left\{e^{t L_t} \right\},\, t\!\in\!\mathbb{R}$, which can be computed using the properties of the compound Poisson distribution. Recall that the \ac{RV} $L_t$ is equal to the sum of  $N\!\sim\!{\text{Po}}\!\left(4\lambda r\right)$ \ac{i.i.d.} uniform \acp{RV} $L_i\!\equiv\!L$ in $\left(0,2r\right)$ plus the constant $4r$. Therefore, 
\begin{equation}
\label{eq:MLt}
\begin{array}{ccl}
\mathbb{E}\!\left\{e^{t L_t} \right\}  &=&  \mathbb{E}_N\!\left\{ \mathbb{E}\!\left\{e^{t L_t} \vert N \right\} \right\} \\ {} &=& \mathbb{E}_N\!\left\{ e^{4rt} M_L\!\left(t\right)^N \right\} \\ {} &\stackrel{(a)}{=}& e^{4rt} \, \mathbb{E}_N\!\left\{ \left( \frac{e^{2tr}-1}{2tr}\right)^N\right\} \\ {} &\stackrel{(b)}{=}& \exp\!\left( 4rt + 4\lambda r \left( \frac{e^{2rt}-1}{2r t} - 1\right) \right), 
\end{array}
\end{equation}
where $(a)$ follows from the \ac{MGF} of a uniform \ac{RV}, and $(b)$ uses the probability generating function of a Poisson \ac{RV}. 

The limit of the first derivative of $M_{L_t}\!\left(t\right)$ with respect to $t$ at $t\!\rightarrow\! 0$ in equation~\eqref{eq:MLt} yields $\left(4r\!+\! 4\lambda r^2\right)$, which is the mean of the \ac{RV} $L_t$. The first term, $4r$, is the fixed length of the typical lines in $B$. The second term, $4\lambda r^2$ is, as expected, equal to the product of the mean length $r$ of a randomly selected line segment $L_i$ multiplied by the expected number $4\lambda r$ of line segments in $B$. Recall that the expected number of line segments is equal to the expected number or intersections along the typical lines. 

Conditionally on the realization of the length $L_t=l$, the number of facilities in $B$ is Poisson distributed with parameter $\lambda_g l$. As a result, using the law of total expectation, the \ac{MGF} of the (discrete) \ac{RV} of the number of facilities in $B$, $N_p\!\left(\Phi_L\cap B\right)$, can be read as 
\[
\begin{array}{ccl}
M_{N_p\left(\Phi_L\cap B\right)}\!\left(t\right) &=& \mathbb{E}\!\left\{e^{N_p\left(\Phi_L\cap B\right) t}\right\} \\ &=& \mathbb{E}_{L_t}\!\left\{ \mathbb{E}\!\left\{e^{N_p\left(\Phi_L\cap B\right) t}\vert L_t\right\} \right\} \\ &\stackrel{(a)}{=}& \mathbb{E}_{L_t}\!\left\{e^{\lambda_g l \left(e^t -1\right) } \right\}  \\ &=& M_{L_t}\!\left(\lambda_g\left(e^t\!-\!1\right)\right), 
\end{array}
\]
where $(a)$ is due to the \ac{MGF} of a Poisson \ac{RV}. 

After substituting the above argument, $\lambda_g\left(e^t\!-\!1\right)$, into the last equality in~\eqref{eq:MLt}, we obtain the \ac{MGF} of the \ac{RV} $N_p\!\left(\Phi_L\cap B\right)$. $M_{N_p\left(\Phi_L\cap B\right)}\!\left(t\right) =$
\[
\exp\!\left(4r\lambda_g\left(e^t\!-\!1\right)\!+\!4\lambda r \left(\frac{e^{2r\lambda_g\left(e^t-1\right)}-1}{2r\lambda_g\left(e^t\!-\!1\right)}-1 \right) \right). 
\]
 
Furthermore, starting from the definition of the \ac{MGF} of a discrete \ac{RV} on the natural numbers, $M_{N_p\left(\Phi_L\cap B\right)}\!\left(t\right)\!\equiv\! M_{N_p}\!\left(t\right)=\sum\nolimits_{k=0}^\infty P_k \, e^{k t}$ we get $P_k$
\begin{equation}
\label{eq:MNp}
\begin{array}{ccl}
{} \!\!\!&=&\!\!\! \left. \frac{1}{k!} \, \frac{{\rm d}^k}{{\rm d}t^k }  M_{N_p}\!\left(\log t\right) \right\vert_{t\rightarrow 0} \\ 
\!\!\!&=&\!\!\!  \left. \frac{1}{k!} \, \frac{{\rm d}^k}{{\rm d}t^k }    \exp\!\left(4r\lambda_g\left(t\!-\!1\right)\!+\!4\lambda r \left(\frac{e^{2r\lambda_g\left(t-1\right)}-1}{2r\lambda_g\left(t-1\right)}\!-\!1 \right) \right) \right \vert_{t\rightarrow 0}. 
\end{array}
\end{equation}

After substituting $k\!=\!0$ in equation~\eqref{eq:MNp}, we obtain $P_0\!=\!\exp\!\left(-4r\lambda_g\!-\!4\lambda r\left(1\!-\!a_0\right) \right)$, as expected, see equation~\eqref{eq:R1}. For $k\!=\!1$ in~\eqref{eq:MNp}, after some simplification, we have
\begin{equation}
\label{eq:Pk1}
P_1 = P_0 \left(4r\lambda_g+4r\lambda \left( a_0-e^{-2r\lambda_g}\right)\right).
\end{equation}

The calculation of higher-order derivatives in~\eqref{eq:MNp} results in complicated expressions which are difficult to manipulate. For instance, we list below the expressions we get, after some simplification, for $P_2$ and $P_3$. 
\begin{equation}
\label{eq:Pk23}
\begin{array}{ccl}
P_2 \!\!\!&=&\!\!\! \frac{1}{2} P_0\left( 4r\lambda_g+4r\lambda \left( a_0-e^{-2r\lambda_g}\right) \right)^2 \!+ 4\lambda r P_0 \times \\ & &  \left(a_0-e^{-2r\lambda_g} -r\lambda_g e^{-2r\lambda_g}\right). \\ 
P_3 \!\!\!&=&\!\!\! \frac{1}{6}P_0 \left( 4r\lambda_g+4r\lambda \left( a_0-e^{-2r\lambda_g}\right) \right)^3 \!+ 4 \lambda r P_0 \times \\ & &  \left(4r\lambda_g + 4r\lambda\left(a_0-e^{-2r\lambda_g} \right)\right) \times \\ \!\!\!& &\!\!\! \left(a_0-e^{-2r\lambda_g} - r\lambda_g e^{-2r\lambda_g}\right) + 4r\lambda P_0 \times \\ & &  \left(a_0-e^{-2r\lambda_g}-r\lambda_g e^{-2r\lambda_g}-\frac{2}{3}r^2\lambda_g^2 e^{-2r\lambda_g}\right)\!.
\end{array}
\end{equation}

One way to add some structure in the calculation of $P_k$, is to use the Fa{\`a} di Bruno's formula, see for instance~\cite{Johnson2002}, for the calculation of the $k$-th derivative of a composite function. Let us define $f\!\left(t\right)\!=\!e^t$ and $g\!\left(t\right)\!=\!\left(4r\lambda_g\left(t\!-\!1\right)\!+\!4\lambda r \left(\frac{e^{2r\lambda_g\left(t-1\right)}-1}{2r\lambda_g\left(t-1\right)}-1 \right)\right)$, see equation~\eqref{eq:MNp}. Leveraging on that $f^{\left(k\right)}\!\left(t\right)\!=\!e^t$, where $f^{\left(k\right)}$ denotes the $k$-th derivative, the Fa{\`a} di Bruno's formula is simplified to~\cite[Eq.~(2.2)]{Johnson2002}: $\frac{{\rm d}^kf\left(g\left(t\right)\right)}{{\rm d}t^k}$
\begin{equation}
\label{eq:Faa}
\begin{array}{ccl}
{} \!\!\! &=& \!\!\! e^{g\left(t\right)} \!\! \sum\limits_{m=1}^k B_{k,m}\!\left(g'\!\left(t\right),g''\!\left(t\right),\ldots g^{\left(k-m+1\right)}\!\left(t\right)\right) \\ {} \!\!\! &=& \!\!\! e^{g\left(t\right)} B_k\!\left(g'\!\left(t\right),g''\!\left(t\right),\ldots g^{\left(k\right)}\!\left(t\right)\right),
\end{array}
\end{equation}
where $B_{k,m}$ and $B_k$ are the partial and the complete exponential Bell polynomials respectively. 

The Bell polynomials emerge in set partitions. For instance, $B_{4,2}\!\left(x_1,x_2,x_3,x_4\right)\!=\!3x_2^2+4x_1 x_3$, indicates that there are three ways to separate the set $\left\{x_1,x_2,x_3,x_4\right\}$ into two subsets of size two, and four ways to separate it into a block of size three and another of size one. Note that the total number of partitions, i.e., seven,  is the Stirling number of second kind which, in general, counts the ways to separate an $m$-element set into $k$ disjoint and non-empty subsets, e.g., $S\!\left(4,2\right)\!=\!7$. The calculation of the Bell polynomials is widely available in today's numerical software packages like Mathematica. 

Recall from equation~\eqref{eq:MNp} that the $k$-th derivative of the \ac{MGF} has to be evaluated in the limit $t\!\rightarrow\!0$. It is also noted that $P_0\!=\! e^{g\left(0\right)}$. Combining equations~\eqref{eq:MNp} and~\eqref{eq:Faa} yields 
\begin{equation}
\label{eq:Faa2}
P_k = \frac{P_0}{k!} \sum\limits_{m=1}^k B_{k,m}\!\left(g'\!\left(0\right),g''\!\left(0\right),\ldots g^{\left(k-m+1\right)}\!\left(0\right)\right).
\end{equation}

Equation~\eqref{eq:Faa2} above is insightful to understand why the calculation of $P_3$ in equation~\eqref{eq:Pk23} consists of three terms. The first term over there, $\left( 4r\lambda_g+4r\lambda \left( a_0-e^{-2r\lambda_g}\right) \right)^3$, is essentially equal to the partial Bell polynomial $B_{3,3}\!\left(g'\!\left(0\right)\right) \!=\!  g'\!\left(0\right)^3$. It is also straightforward to verify that the remaining two terms in~\eqref{eq:Pk23} are equal to $B_{3,2}\!\left(g'\!\left(0\right), g''\!\left(0\right)\right) \!=\! 3 g'\!\left(0\right)  g^{\left(2\right)}\!\left(0\right)$ and $B_{3,1}\!\left(g'\!\left(0\right), g''\!\left(0\right),g^{\left(3\right)}\!\left(0\right)\right) \!=\!  g^{\left(3\right)}\!\left(0\right)$.

The Fa{\`a} di Bruno's formula can indeed provide some insight into the calculation of the \ac{MGF} of the \ac{RV} describing the total number of points in $B$, $N_p\!\left(\Phi_L\cap B\right)$, however, the cost of computing the $k$-order partial derivatives might be high, especially for large values of $k$. In the next section, we will use enumerative combinatorics, revealing a simple numerical algorithm to evaluate $P_k$ without involving higher-order derivatives as in equation~\eqref{eq:Faa2}. 
 
\section{Calculating $P_k$ using integer partitions}
\label{sec:Integer}
Let us assume there are $k$ facilities in $B$ and separate their allocation into two sets: along the typical segments $L_x, L_y$ and in the rest of $B$. The probability $P_k = \mathbb{P}\!\left(N_p\!\left(\Phi_L\cap B\right) = k \right)$ can be read as $P_k =$
\begin{equation}
\label{eq:Pk}
\sum\limits_{i\leq k} \mathbb{P}\!\left(N_p\!\left(\left\{L_x\cup L_y \right\}\cap B\right) = i \right) \cdot \mathbb{P}\!\left(N_p\!\left(\Phi_l \cap B\right) = m \right),  
\end{equation}
where $\left(m\!=\!k\!-\!i\right)$, and the product of probabilities follows from the independent locations of intersections along $L_x$ and $L_y$, and the independent locations of facilities along each line of $\Phi_L$. 

The first probability term in~\eqref{eq:Pk} can be calculated as $\mathbb{P}\!\left(N_p \!\left(\left\{L_x\cup L_y \right\}\cap B\right) = i \right)$
\begin{equation}
\label{eq:Pka}
\begin{array}{lll}
 {} \!\!\!&=&\!\!\! \mathbb{P}\!\left(N_p\!\left(L_x\cap B\right) + N_p\left( L_y \cap B\right) = i \right) \\ 
{} \!\!\!&\stackrel{(a)}{=}& \frac{\left( 4\lambda_g r\right)^i e^{-4\lambda_g r}}{i!}, 
\end{array}
\end{equation}
where $(a)$ uses the fact that the superimposed \acp{PPP} of facilities along $L_x$ and $L_y$ is another \ac{PPP} with twice the intensity $2\lambda_g$. 

The calculation of the second probability term in~\eqref{eq:Pk} is more involved, but similar to~\eqref{eq:Pka}, it helps to consider just a single \ac{PPP} of intersection points with twice the intensity, $2\lambda$, along $L_x$ instead of two line processes $\Phi_{lh}, \Phi_{lv}$. Let us denote the resulting distribution of vertical lines by $\Phi_{lv}'$. Obviously, $\mathbb{P}\!\left(N_p\!\left(\Phi_l \cap B\right) = m \right)\!=\!\mathbb{P}\!\left(N_p\!\left(\left\{\Phi_{lv},\Phi_{lh}\right\} \cap B\right) = m \right)\!=\! \mathbb{P}\!\left(N_p\!\left(\Phi_{lv}' \cap B\right) = m \right)$. The latter can be written as $\mathbb{P}\!\left(N_p\!\left(\Phi_{lv}' \cap B\right) = m \right)$
\begin{equation}
\label{eq:Pk2}
\begin{array}{ccl}
{} \!\!\!&=&\!\!\!  
\sum\limits_{n=0}^\infty  \mathbb{P}\!\left( N_p\left(\Phi_{lv}'\cap B \right)\!=\!m \vert  N \!=\! n\right) \cdot \mathbb{P}\!\left( N \!=\! n\right)\\ \!\!\!&=&\!\!\!
\sum\limits_{n=0}^\infty \frac{\left(4\lambda r\right)^n e^{-4\lambda r}}{n!} \mathbb{P}\!\left( N_p\left(\Phi_{lv}'\cap B \right)=m \vert  N = n\right). 
\end{array}
\end{equation}
 
In order to calculate the conditional probability in~\eqref{eq:Pk2}, we have to enumerate the number of ways of allocating $m$ facilities (or objects) into $n$ lines (or urns), with both objects and urns being indistinct. For each possible allocation, we need to obtain its probability of occurrence, and finally we will sum over all obtained values. For $n\!\geq\! m$, the number of ways to allocate $m$ objects into $n$ urns is equal to the number of integer partitions of $m$, denoted by $p\left(m\right)$, because only the number of objects going to each urn is relevant. For $n\!<\!m$, the restricted integer partitions of size at most $n$, $p_n\!\left(m\right)$, have to be considered. Empty urns are obviously allowed. Next, we sum over the probabilities of all partitions. $\mathbb{P}\!\left( N_p\left(\Phi_{lv}'\cap B \right)=m \vert  N = n\right) =$
\[
\sum\limits_{\xi\in p_n\!\left(m\right)} \mathbb{P}\!\left( N_p\left(\Phi_{lv}'\cap B \right)=m \vert  N = n,\xi \right),
\]
where $p_n\!\left(m\right)\!=\!p\left(m\right)$ for $n\!\geq\! m$ and $\xi$ is the set associated with a partition, e.g.,  $p\left(3\right)\!=\!\left\{\left\{3\right\}, \left\{2,1\right\}, \left\{1,1,1\right\}\right\}$. 

After substituting the above equality in the last line of~\eqref{eq:Pk2}, and interchanging the orders of summations we end up with $\mathbb{P}\!\left(N_p\!\left(\Phi_{lv}' \cap B\right) \!=\! m \right) =$
\begin{equation}
\label{eq:Pk3}
\sum\limits_{\xi\in p\left(m\right)}\sum\limits_{n=\left|\xi\right|}^\infty \!\!  \frac{\left(4\lambda r\right)^n e^{-4\lambda r}}{n!} \mathbb{P}\!\left( N_p\left(\Phi_{lv}'\cap B \right) \!=\! m \vert  N \!=\! n,\xi\right), 
\end{equation}
where $\left|\cdot \right|$ denotes the set cardinality, e.g., $\left|\xi\right|\!=\!2$ for $\xi\!=\!\left\{2,1\right\}$. 

At this point, it helps to define the parameter $a_k, k\in\mathbb{N}$ describing the probability that a vertical line with abscissa $z>0$, uniformly distributed between the origin and the point $\left(r,0\right)$, contains exactly $k$ facilities in $B$, e.g., in  Fig.~\ref{fig:Grids}, the line with abscissa $z_1$ does not contain any.  
\begin{equation}
\label{eq:ak}
\begin{array}{ccl}
a_k &=& \displaystyle \frac{1}{r} \int_0^r \frac{\left(2\lambda_g\left(r-z\right)\right)^k}{\Gamma\!\left(k\!+\!1\right)} \,  e^{-2\lambda_g \left(r-z\right)} {\rm d}z \\ &=& \frac{\Gamma\left(k+1,2\lambda_g r\right)}{2\lambda_g r},
\end{array}
\end{equation}
where $\Gamma\!\left(k\!+\!1\right)=k!$, $\Gamma\!\left(\alpha,x\right)=\frac{1}{\Gamma\left(\alpha\right)} \int_0^x t^{\alpha-1} e^{-t} {\rm d}t$ is the lower incomplete Gamma function, and for $k\!=\!0$ we obtain the parameter $a_0$ defined under equation~\eqref{eq:R1}.

Let us consider the partition $\xi \!=\! \left\{1,1,\ldots, 1\right\}$ with $m\,  {\text{1's}}$. The inner sum in equation~\eqref{eq:Pk3}, conditionally on this partition, yields $\mathbb{P}\!\left(N_p\!\left(\Phi_{lv}' \cap B\right) = m \vert \xi \right)$
\[
\begin{array}{ccl}
{} \!\!\!&=&\!\!\!  
\sum\limits_{n=m}^\infty \frac{\left(4\lambda r\right)^n e^{-4\lambda r}}{n!} \cdot \mathbb{P}\!\left( N_p\left(\Phi_{lv}'\cap B \right)=m \vert  N = n,\xi \right) \\ \!\!\!&=&\!\!\! \sum\limits_{n=m}^\infty \frac{\left(4\lambda r\right)^n e^{-4\lambda r}}{n!} \cdot \binom{n}{m} a_1^m a_0^{n-m} \\ \!\!\!&=&\!\!\! \frac{\left(4\lambda r\right)^m a_1^m e^{-4\lambda r \left(1-a_0\right)}}{m!},
\end{array}
\]
where the binomial coefficient $\binom{n}{m}$ represents the number of ways to select the $m$ lines containing just one facility in $B$. 

Substituting the above equation and equation~\eqref{eq:Pka} into~\eqref{eq:Pk} yields the conditional probability, $P_{k\vert\xi}$, of $k$ facilities in $\left\{\Phi_{lv}'\cup L_x\right\}\cap B$ given the partition $\xi$. 
\[
\begin{array}{ccl}
P_{k\vert \xi} \!\!\!&=&\!\!\! \sum\limits_{i=0}^k \frac{\left( 4\lambda_g r\right)^i e^{-4\lambda_g r}}{i!} \frac{\left(4\lambda r\right)^{k-i} a_1^{k-i} e^{-4\lambda r \left(1-a_0\right)}}{\left(k-i\right)!} \\ \!\!\!&=&\!\!\! \frac{\left(4 r\left(\lambda_g+\lambda a_1\right)\right)^k \, P_0}{k!}.
\end{array}
\]

In a similar manner, we can evaluate $P_{k\vert \xi}$ for all $\xi\!\in\!p\!\left(k\right)$ and complete the calculation of $P_k\!=\!\sum\nolimits_{\xi\in p\left(k\right)} \!\! P_{k\vert \xi}$ in~\eqref{eq:Pk}. However, this might be cumbersome, unless a simple pattern is identified. Next, we will derive a simple expression for $P_{k\vert\xi}$, depending on the number of times an integer appears in the partition $\xi$. Let us consider $\xi\!=\!\left\{q,\ldots, q, 1,\ldots ,1\right\}$, where the $q$ appears $f$ times and there are also $\left(m\!-\!q f\right)$ 1's. The inner sum in equation~\eqref{eq:Pk3}, conditionally on the partition $\xi$ with cardinality $\left|\xi\right|\!=\!\left(\left(m\!-\!q f\right)\!+\!f\right)$, yields $\mathbb{P}\!\left(N_p\!\left(\Phi_{lv}' \cap B\right)  = m \vert \xi \right)$
\[
\begin{array}{ccl}
{} \!\!\! &=& \!\!\!
\sum\limits_{n=m-\left(q-1\right)f}^\infty \frac{\left(4\lambda r\right)^n e^{-4\lambda r}}{n!} \,  a_q^f \, a_1^{m-q f} \,  a_0^{n-\left(m-\left(q-1\right)f\right)} \times \\ & & \,\,\,\,\,\,\,\,\, \binom{n}{m-\left(q-1\right)f} \binom{m-\left(q-1\right)f}{f} \\  \!\!\! &=& \!\!\! \binom{m-\left(q-1\right)f}{f} a_q^f \, a_1^{m-q f}\, \frac{\left(4\lambda r\right)^{m-\left(q-1 \right)f}e^{-4\lambda r\left(1-a_0\right)}}{\left(m-\left(q-1 \right)f\right)!},  
\end{array}
\]
where $\left|\xi\right|\!=\!\left(m\!-\!\left(q\!-\!1\right)f\right)$ is the number of lines containing facilities in $B$. They are selected with $\binom{n}{m-\left(q-1\right)f}$ ways from the available $n$ lines, and $\binom{m-\left(q-1\right)f}{f}$ is the number of ways to select $f$ out of the segments containing facilities in $B$, and allocate $q$ facilities to each one of them.

Keeping in mind that due to the existence of $f$ replicas of the integer $q$ in the partition, only up to $\left(k-qf\right)$ facilities may be located along $L_x$, we substitute the above equation and~\eqref{eq:Pka} into~\eqref{eq:Pk}, ending up with 
\begin{equation}
\label{eq:Pkpartition}
\begin{array}{ccl}
P_{k\vert \xi} \!\!\!\!\!&=&\!\!\!\!\! \sum\limits_{i=0}^{k-qf} \frac{\left( 4\lambda_g r\right)^i e^{-4\lambda_g r}}{i!} \binom{k-i-\left(q-1\right)f}{f} a_q^f \, a_1^{k-i-q f}\, \times \\ & & \,\,\, \frac{\left(4\lambda r\right)^{k-i-\left(q-1 \right)f}e^{-4\lambda r\left(1-a_0\right)}}{\left(k-i-\left(q-1 \right)f\right)!} \\ \!\!\!&=&\!\!\! \frac{\left(4r\left(\lambda_g+\lambda a_1\right)\right)^{k-fq}}{\left(k-fq\right)!} \, \frac{\left(4\lambda r a_q\right)^f}{f!} P_0.
\end{array}
\end{equation}

It is straightforward to generalize the above calculation to include partitions with more than one $q\!>\!1$. Let us assume that the integer $q_i\!>\!1$ appears $f_i\!\geq\!1$ times in the partition $\xi$. Equation~\eqref{eq:Pkpartition} can be generalized as  
\begin{equation}
\label{eq:Pkpartition2}
P_{k\vert \xi} =  \frac{\left(4r\left(\lambda_g\!+\!\lambda a_1\right)\right)^{k-\sum\nolimits_i f_i q_i}\, P_0}{\left(k-\sum\nolimits_i f_i q_i \right)!} \cdot \prod\nolimits_{q_i} \!\!\! \frac{\left(4\lambda r a_{q_i}\right)^{f_i}}{f_i!}.
\end{equation}

To sum up, in order to evaluate $P_{k\vert\xi}$, we start with $P_{k\vert\xi}\!=\!P_0$. Given that the integer $q\!>\!1$ appears in the partition $f\!\geq\! 1$ times, we set $P_{k\vert\xi}\leftarrow P_{k\vert\xi} \frac{\left(4\lambda r a_q\right)^f}{f!}$. For $q\!=\!1$, we set  $P_{k\vert\xi}\leftarrow P_{k\vert\xi} \frac{\left(4r\left(\lambda_g+\lambda a_1\right)\right)^f}{f!}$. We update $P_{k\vert\xi}$ for all integers  $q\!\in\!\xi$. Next, we repeat the same procedure for all partitions $\xi$, and we compute the probability $P_k\!=\!\sum\nolimits_{\xi\in p\left(k\right)} P_{k\vert\xi}$. 

For illustration purposes, in Table~\ref{table:table1}, we list the contributions of the seven different terms involved in the calculation of $P_5$. The inputs in the rightmost column, which is equal to the product of the terms in the middle column, are generated based on equation~\eqref{eq:Pkpartition2}. For instance, for the partition $\left\{3,2\right\}$ we have $f_1\!=\!f_2\!=\!1$, because each of the numbers $q_1=2,q_2=3$ appears only once in the partition. Furthermore, $\left(f_1q_1+f_2q_2=5\right)$ and thus, the exponent of the term $\left(\lambda_g\!+\!\lambda a_1\right)$ is zero. Therefore, equation~\eqref{eq:Pkpartition2} degenerates to the product of just two terms, $4\lambda r a_2$ and $4\lambda r a_3$, scaled by $P_0$, and the result for the probability $P_{5\vert\left\{3,2\right\}}$ directly follows. 

Now, it becomes clear in the calculation of $P_3$ in equation~\eqref{eq:Pk23} that the first term corresponds to the partition $\left\{1,1,1\right\}$ with $a_1\!=\!a_0\!-\!e^{-2r\lambda_g}$, the second term to the partition  $\left\{2,1\right\}$ with $a_2\!=\!a_0-e^{-2r\lambda_g} - r\lambda_g e^{-2r\lambda_g}$, and the last term to the partition $\left\{3\right\}$ with $a_3\!=\!\frac{1}{2}\left(a_0\!-\!e^{-2r\lambda_g}\!-\!r\lambda_g e^{-2r\lambda_g}\!-\!\frac{2}{3}r^2\lambda_g^2 e^{-2r\lambda_g} \right)$. For completeness, the pseudocode used to calculate $P_k$ is provided as Algorithm~1. Given $P_j, \, \forall j\!\in\!\left\{0,1, \ldots\left(k\!-\!1\right)\right\}$, the \ac{CDF} of the distance distribution is evaluated as $F_{R_k}\!\left(r\right)\!=\! 1 \!-\! \sum\nolimits_{j=0}^{k-1} P_j$. 
\begin{table}[!t]
\caption{Detailing the calculation of $P_5$ using integer partitions. See also equation~\eqref{eq:Pkpartition2}.}
\label{table:table1}
\centering
\tiny
\setlength\extrarowheight{5pt}
\begin{tabular}{|c|c|c|} 
 \hline
  Partition & Terms & Probability $P_{k\vert\xi}$  \\ \hline  
 $\left\{5\right\}$ & $4\lambda r a_5$ & $4\lambda r a_5 P_0$ \\  \hline 
 $\left\{4,1\right\}$ & $4 \lambda r a_4, 4 r \left(\lambda_g\!+\!\lambda a_1\right)$ & $16\lambda \left(\lambda_g\!+\!\lambda a_1\right) r^2 a_4 P_0 $  \\  \hline $\left\{3,2\right\}$ & $4 \lambda r a_3, 4\lambda r a_2$ & $16\lambda^2 r^2 a_2 a_3 P_0$  \\  \hline
$\left\{3,1,1\right\}$ & $4 \lambda r a_3, \frac{1}{2} \left(4r\left(\lambda_g\!+\!\lambda a_1\right)\right)^2$ & $32\lambda \left(\lambda_g\!+\!\lambda a_1\right)^2 r^3 a_3 P_0$ \\  \hline $\left\{2,2,1\right\}$ & $\frac{1}{2} \left(4r\lambda a_2\right)^2, 4 r \left(\lambda_g\!+\!\lambda a_1\right)$ & $32\lambda^2 \left(\lambda_g\!+\!\lambda a_1\right) r^3 a_2^2 P_0$ \\ \hline $\left\{2,1,1,1\right\}$ & $4r\lambda a_2, \frac{1}{6} \left(4 r \left(\lambda_g\!+\!\lambda a_1\right)\right)^3$ & $\frac{128}{3} \lambda \left(\lambda_g\!+\!\lambda a_1\right)^3 r^4 a_2 P_0$ \\[5pt]  \hline $\left\{1,1,1,1,1\right\}$ & $\frac{1}{120} \left(4 r \left(\lambda_g\!+\!\lambda a_1\right)\right)^5$ & $\frac{128}{15} \left(\lambda_g\!+\!\lambda a_1\right)^5 r^5 P_0$ \\ \hline
\end{tabular}
\end{table}
\begin{algorithm}
\caption{Compute $P_k$}
\begin{algorithmic}[1]
\STATE $a_q\gets \frac{\Gamma\!\left(q+1,2\lambda_g r\right)}{2\lambda_g r}, \, q=0,1,\ldots \left(k\!-\!1\right)$
\STATE $P_0\gets e^{-4\lambda_g r -4\lambda r\left(1-a_0\right)}, \, P_k \gets 0$
\STATE $\Xi={\text{IntegerPartitions}}\left(k\right)$
\FORALL {$\xi\in\Xi$}
\STATE $P_{k\vert \xi}\gets P_0$
\FORALL {$q\in\xi$}
\STATE $f_q\gets {\text{card}}_\xi\! \left(q\right)$ 
\IF {$q=1$}
\STATE $P_{k\vert \xi}\gets P_{k\vert \xi} \frac{\left(4\left(\lambda_g+\lambda a_1\right)r\right)^{f_q}}{f_q!}$
\ELSE
\STATE $P_{k\vert \xi}\gets P_{k\vert \xi} \frac{\left(4\lambda a_q r\right)^{f_q}}{f_q!}$
\ENDIF
\ENDFOR
\STATE $P_k\gets P_k + P_{k\vert \xi}$
\ENDFOR
\end{algorithmic}
\end{algorithm}

Thus far, we have considered equal intensity of intersections $\lambda_v=\lambda_h=\lambda$ along the typical lines $L_x, L_y$. The calculations in Algorithm~1 can be easily generalized to $\lambda_v\neq\lambda_h$. Specifically, one will have to use  
\begin{equation}
P_0\gets e^{-4\lambda_g r -2\left(\lambda_v+\lambda_h\right) r\left(1-a_0\right)}, 
\end{equation}
in line 2, and
\begin{equation}
\label{eq:DiffIntens}
\begin{array}{ccl} 
P_{k\vert \xi} &\gets& P_{k\vert \xi} \frac{\left(2\left(2\lambda_g+\left(\lambda_v+\lambda_h\right) a_1\right)r\right)^{f_q}}{f_q!} \\
P_{k\vert \xi} &\gets& P_{k\vert \xi} \frac{\left(2\left(\lambda_v + \lambda_h\right) a_q r\right)^{f_q}}{f_q!}, 
\end{array}
\end{equation}
in lines 9 and 11 respectively. Also, the intensity of intersections for the full Manhattan grid can be read as $\lambda=\frac{2\lambda_v \lambda_h}{\lambda_v+\lambda_h}$.

\section{Numerical illustrations \& Applications}
\label{sec:Numericals}
In this section, we will start by validating Algorithm~1 with simulations, and we will proceed with the study of three example scenarios where this Algorithm might be of use. Finally, we will show that Algorithm~1 can give a short and accurate estimate about the distance distributions in real road networks with an approximate regular street layout. For that, we will use the map of an area near Manhattan in New York with the map data retrieved using OpenStreetMap~\cite{OSM, Filippidis}.    

\begin{figure}[!t]
 \centering
\includegraphics[width=3in]{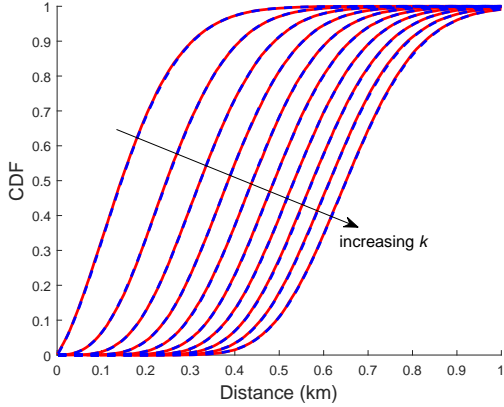}\label{fig:PathsCDF2a}
 \caption{The distance distributions for the $k$ nearest facilities to the origin $k\!\in\!\left\{1,2,\ldots 10\right\}$ following a \ac{MPLCP} with dense streets $\lambda\!=\!10 \, {\text{km}}^{-1}$ and sparse facilities $\lambda_g\!=\!0.5 \, {\text{km}}^{-1}$. The red lines are averages over $50\, 000$ simulations and the dashed blue lines are (exact) calculations using Algorithm 1. The simulations are carried out within a square area of $400\, {\text{km}}^2$. The typical intersection (origin) is placed at the middle of the square. In each simulation, an independent realization of the \ac{MPLCP} is generated and the path distances from the origin to the $k$ nearest points of the process are computed. Note that in the calculation of $P_9$, Algorithm 1 should go through $30$ partitions but the computational complexity still remains very low.}
 \label{fig:PathsCDF2}
\end{figure}

\begin{figure}[!t]
 \centering
\includegraphics[width=3in]{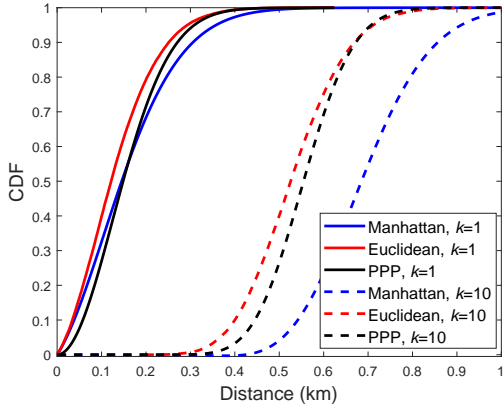}
\caption{The $k$-th nearest neighbor distance distributions $k\!\in\!\left\{1,10\right\}$ from a random intersection, for a \ac{MPLCP} with $\lambda\!=\!10 \, {\text{km}}^{-1}$ and $\lambda_g\!=\!0.5 \, {\text{km}}^{-1}$, using the Manhattan and the Euclidean distance. The associated distance distributions for a planar \ac{PPP} of equal intensity are also demonstrated. The curves are generated using Algorithm 1 for the Manhattan distance, simulations for the Euclidean distance and equation~\eqref{eq:PPPRn} for the \ac{PPP}.}
\label{fig:L1L2PPP}
\end{figure}

In Fig.~\ref{fig:PathsCDF2} we have validated the calculation of the path distance distribution using Algorithm~1 for the ten nearest neighbors of a \ac{MPLCP} $k\!\leq\!10$. Fig.~\ref{fig:L1L2PPP} illustrates that the Euclidean distance (L2 norm) is a bad approximation to the Manhattan distance (L1 norm) distribution for a \ac{MPLCP}. The approximation quality deteriorates for a larger $k$. The planar \ac{PPP} of equal intensity, $\mu\!=\!2\lambda\lambda_g$, where the locations of facilities are not constrained by the road network, is not a better approximation either. Note that for the \ac{PPP}, the distance to the $k$-th nearest neighbor follows the generalized gamma distribution with \ac{PDF}~\cite[Theorem 1]{Haenggi2005}: 
\begin{equation}
\label{eq:PPPRn}
f_{R_k}\!\left(r\right) = \frac{2 \, e^{-\mu\pi r^2}\left(\mu\pi r^2\right)^k}{r\, \Gamma\!\left(k\right)}.
\end{equation}

Having justified that the planar \ac{PPP} and the Euclidean distances are not good approximations to the path distances, we will  next present some case studies where the path distance distributions can be of use. 

\subsection{Distance distributions in spatial database queries}
Let us consider an electric vehicle at an intersection querying for the nearest charging station. The charging stations might be closed or fully occupied, depending on the time of day and the road traffic conditions. In that case, the spatial database should respond to the query by returning the {\it{nearest available}} charging station to the vehicle. The distance distributions  developed in this paper can be used to calculate the path distance distribution and the distribution of travel time to the {\it{nearest available}} facility. Given that any facility is available with probability $q$, independently of other facilities, and the average travel speed is $v$, the \ac{CDF} of the average travel time to the {\it{nearest available}} facility follows from the geometric distribution: 
\begin{equation}
\label{eq:Pt}
\mathbb{P}\!\left(t\!\leq\!\tau\right) = \sum\limits_{i=1}^\infty q \left(1-q\right)^{i-1} F_{R_i}\!\left(\tau\right),  
\end{equation}
where $\tau\!=\!r v^{-1}$. 

See Fig.~\ref{fig:TravelTimeQ06} for the validation of~\eqref{eq:Pt}. Note that the underlying assumption in equation~\eqref{eq:Pt} is that the delay at the intersection and the traffic-related delays are not modeled explicitly but are incorporated into the model through the average velocity $v$.   

\subsection{Planning the network of facilities in a city}
\begin{figure}[!t]
 \centering
\subfloat[$q=0.6$ ]{\includegraphics[width=3in]{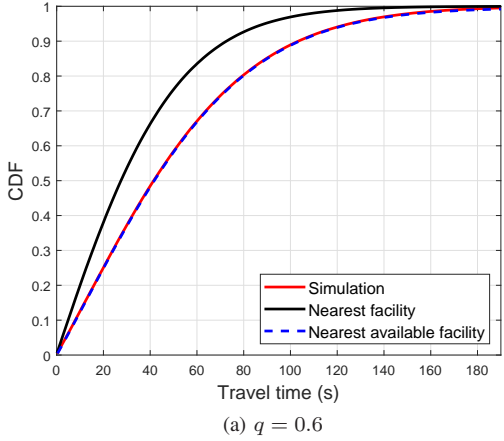} \label{fig:TravelTimeQ06}} \hfil
\subfloat[$q=0.2$ ]{\includegraphics[width=3in]{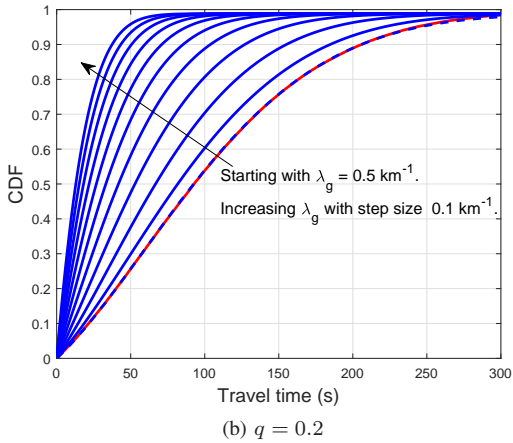} \label{fig:TravelTimeQ02}}
 \caption{Distribution of travel time to the nearest available facility from an intersection. Each facility is available with probability $q$ independently of other facilities. $\lambda\!=\! 1 \, {\text{km}}^{-1}, \, \lambda_g\!=\! 0.5\, {\text{km}}^{-1}, \,  v\!=\! 10 {\text{m/s}}$. The sum in~\eqref{eq:Pt} is truncated at $i\!=\!20$ and validated against the simulations.}
\end{figure}

Before starting to build charging stations for electric vehicles in a city, it is important to identify their required density, i.e., the minimum number of stations per square kilometer, so that certain design constraints are satisfied. This process resembles network dimensioning in wireless communications. Given the intensity of roads $\lambda$, we would like to identify the minimum required intensity of facilities $\lambda_g$ so that a vehicle at a randomly selected intersection can arrive at the nearest available facility within the target time, e.g., $100$ s with probability larger than $90 \%$. Due to the low computational complexity of the path distance distributions using Algorithm~1, we can obtain the required intensity $\lambda_g$ numerically. To give an example, for the parameter settings used to generate Fig.~\ref{fig:TravelTimeQ02}, the above target can be safely met for $\lambda_g\!\geq\!1\, {\text{km}}^{-1}$.

\subsection{Urban vehicular communication networks}
\begin{figure}[!t]
 \centering
\includegraphics[width=3in]{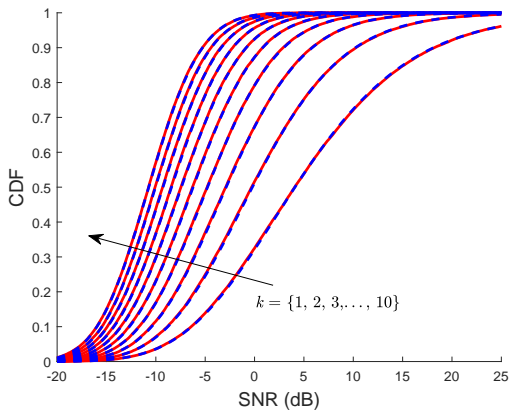}
\caption{The \ac{SNR} distribution at the $k$ nearest vehicles $k\!\leq\!10$ with NLoS connection to the \ac{RSU}. The locations of vehicles follow a \ac{MPLCP} with  $\lambda\!=\!5 \, {\text{km}}^{-1}$ and $\lambda_g\!=\!10 \, {\text{km}}^{-1}$.  Distance-based pathloss $r^{-\eta}$ with $\eta\!=\!3$, diffraction loss around the corner $20$ dB and noise power level $N_0\!=\! 10^{-8}$. The simulations are depicted in `red' and the (exact) calculations in `blue'.}
\label{fig:SNRCox}
\end{figure}

Turning our interest to  wireless communications applications, we assume that a \ac{RSU} is deployed at the typical intersection and the locations of vehicles follow a \ac{MPLCP}. The \ac{RSU} broadcasts messages to the vehicles. For wireless propagation along urban street micro cells, the pathloss model should be different for line- and \ac{NLoS} vehicles~\cite[Fig.~5]{Andersen1995}. The vehicles with \ac{NLoS} connections suffer from serious diffraction losses due to the propagation of wireless signals around the corner. In Fig.~\ref{fig:SNRCox} the distribution of the \ac{SNR} for the $10$ nearest vehicles with \ac{NLoS} connection to the \ac{RSU} is depicted. Assuming a distance-based propagation pathloss $r^{-\eta}$, where $r$ stands for the Manhattan distance, and a diffraction loss $\mathcal{L}$, it is straightforward to convert the distance distributions into received signal level distributions. Then, it also remains to scale the obtained \acp{CDF} by the noise power level $N_0$. Specifically, for the $k$-th nearest vehicle with \ac{NLoS} connection we have   
\begin{equation}
\label{eq:SNRCDF}
\begin{array}{ccl}
\mathbb{P}\!\left({\text{SNR}}_k\leq \theta\right) &=& \mathbb{P}\left(\mathcal{L} R_k^{-\eta}\leq \theta N_0\right) \\ &=& 1- \mathbb{P}\left(R_k\leq \left(\theta N_0/\mathcal{L}\right)^{-1/\eta} \right) \\ &=& \sum\limits_{j=0}^{k-1}\sum\limits_{\xi\in p\left(j\right)} P_{j\vert\xi}\big\vert_{r=\left(\theta N_0/\mathcal{L}\right)^{-1/\eta}}, 
\end{array}
\end{equation}
where the vehicles along $L_x, L_y$ with a line-of-sight connection to the \ac{RSU} are neglected, hence,~\eqref{eq:Pkpartition2} degenerates to  
\[
P_{k\vert \xi} =  e^{-4\lambda r \left(1-a_0\right)}  \cdot \prod\nolimits_{q_i} \!\!\! \frac{\left(4\lambda r a_{q_i}\right)^{f_i}}{f_i!}.
\]

\begin{figure}[!t]
 \centering
\includegraphics[width=3in]{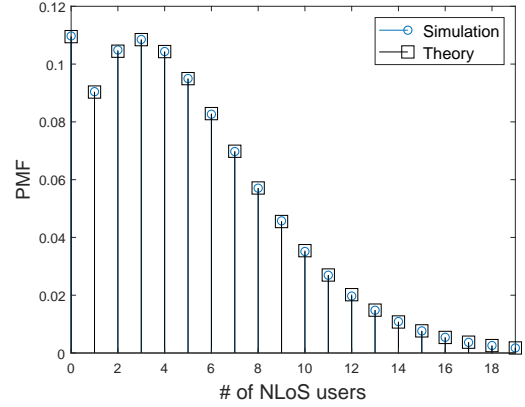}
\caption{The distribution of the number of \ac{NLoS} users within a \ac{RSU} cell whose size is selected to give an \ac{SNR} equal to $-6$ dB at the cell edge. For the rest of the parameter settings see the caption in Fig.~\ref{fig:SNRCox}.}
\label{fig:LoadCoxNLoS}
\end{figure}

Given the size of the cell $B$, see Fig.~\ref{fig:Grids}, it is straigtforward to convert the distance distributions for the \ac{NLoS} vehicles into the distribution of their number inside the cell $-$ network load distribution. Since the \ac{NLoS} vehicles have much lower link gains than the vehicles along the typical lines $L_x, L_y$, the \ac{RSU} must allocate to them more spectral resources under some fair scheduling scheme. Therefore our ability to quickly characterize the network load distribution for NLoS vehicles, see Fig.~\ref{fig:LoadCoxNLoS} for an example illustration, is important.  

\subsection{Model validation with a real map}
\begin{figure}[!t]
 \centering
\subfloat[https://www.openstreetmap.org/]{\includegraphics[width=3in]{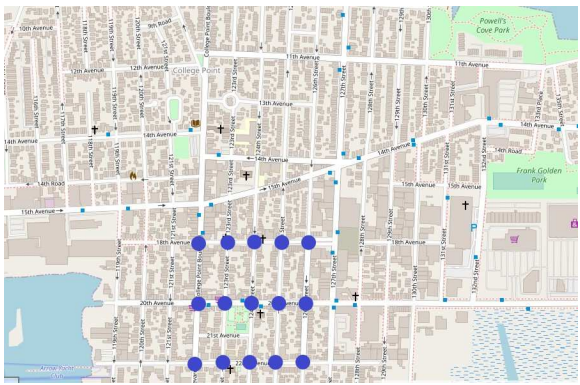} \label{fig:MapNY}} \hfil
\subfloat[https://www.mathworks.com/products/automated-driving.html]{\includegraphics[width=3in]{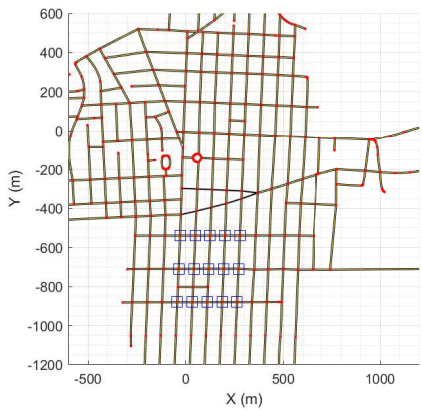}\label{fig:RoadNY}}
\caption{(a) The map of an area near Manhattan, New York (40.7841/-73.8430). (b) The  associated road network is extracted using Matlab toolboxes. All two-dimensional coordinates representing the start and the end points of road segments, generated by parsing the OpenStreetMap file, are depicted in red dots. The $15$ intersection points where the shortest paths are simulated are depicted as blue squares; these points are illustrated as blue disks in Fig.~\ref{fig:MapNY}. The total length of streets in the considered area is approximately $40\, {\text{km}}$.}
\end{figure}

In this section, we investigate the effectiveness of the proposed approach in a practical setting, where the layout of the road network does not precisely follow the \ac{MPLP}. For this purpose, we have extracted the road network of an urban area of size $1.8\,{\text{km}}^2$ using~\cite{OSM,Filippidis}, and  selected $15$ points, forming approximately a grid, where the distance distributions of the shortest paths are simulated. The map of the area can be seen in Fig.~\ref{fig:MapNY} and the associated road network in Fig.~\ref{fig:RoadNY}. The extracted road network consists of approximately $500$ linear segments. We assume that the locations of facilities follow a one-dimensional \ac{PPP} of intensity $\lambda_g$ along each segment. Therefore the only difference between the model and the studied scenario is the underlying road layout.  

\begin{figure}[!t]
 \centering
\subfloat[$\lambda_g=0.5\, {\text{km}}^{-1}$ ]{\includegraphics[width=3in]{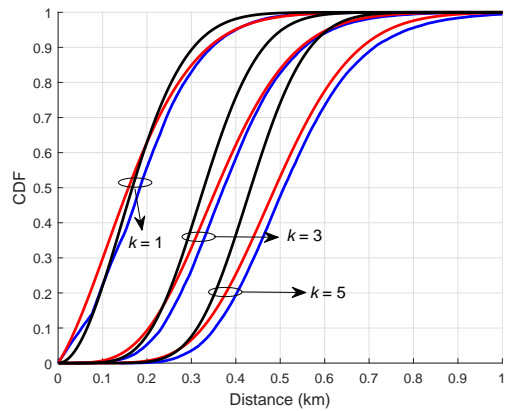}\label{fig:MapLamgA}} \hfil
\subfloat[$\lambda_g=1\, {\text{km}}^{-1}$ ]{\includegraphics[width=3in]{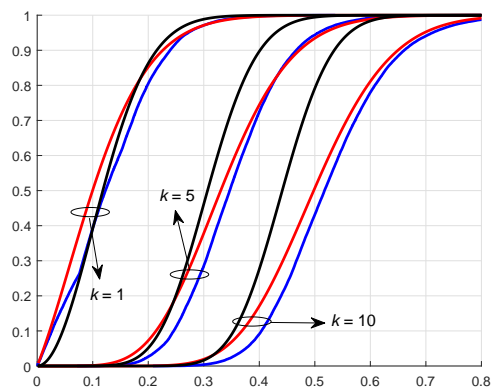}\label{fig:MapLamgB}}
 \caption{Shortest path distance distributions for different intensities of facilities $\lambda_g$ on a real map for $15$ intersection points, see Fig.~\ref{fig:MapNY}. Near the intersection points, the intensity of  horizontal streets is $\lambda_h \approx 5.9 \, {\text{km}}^{-1}$ and the intensity of vertical streets is $\lambda_v \approx 12.5 \, {\text{km}}^{-1}$ yielding $\lambda\!\approx\! 7.9 \, {\text{km}}^{-1}$, which is used in Algorithm~1. 'Blue' lines are simulations, 'red' lines correspond to the \ac{MPLCP} model (Algorithm~1) and 'black' lines to the two-dimensional \ac{PPP} model, see equation~\eqref{eq:PPPRn}. $2\,500$ simulation runs per intersection point. In each run an independent realization of facilities with the specified intensity $\lambda_g$ is generated along every street segment. Then the shortest path distances are calculated using the Dijkstra algorithm.}
\end{figure}

In Fig.~\ref{fig:MapLamgA}, we can see that the performance accuracy of Algorithm~1 improves for larger $k$, while, on the other hand, the prediction accuracy of the two-dimensional \ac{PPP} degrades. In addition, increasing the density of facilities to $\lambda_g\!=\!1\, {\text{km}}^{-1}$, compare Fig.~\ref{fig:MapLamgA} with Fig.~\ref{fig:MapLamgB} for the same value of $k$, is associated with worse model performance. That is because, for denser facilities, the $k$ nearest neighbors are likely to come closer to the intersection points. As a result, the characteristics of the road network near the intersection points, which do not follow exactly the \ac{MPLP} model, start to affect more.    

We conclude this section by evaluating the model performance for a very low density of facilities. In this case, we use only the intersection point at the top-right corner of the grid located near the center of the total area. It can be considered that this point best represents the distance distributions in large urban areas, which, unfortunately, we do not have access to available data. We can see in Fig.~10 that the model performance further improves. The facilities spread out far from the intersection point, hence, the underlying road structure starts to have less effect on the path distance separation. That is a promising finding because it indicates that, for sparse facilities, the suggested model is fairly valid independently of the underlying urban road layout. However, further data collection from different cities is required to ascertain this finding. 
\begin{figure}[!t]
 \centering
\includegraphics[width=3in]{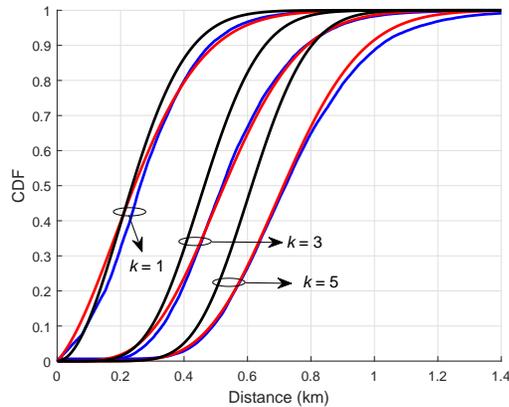}
 \caption{Path distance distributions for the intersection point at the top-right corner of the grid of the $15$ intersection points, see Fig.~\ref{fig:MapNY}. Intensity of facilities $\lambda_g=0.25\, {\text{km}}^{-1}$. For the rest of the parameter settings, see the caption of the previous figure.}
\label{fig:VerySparse}
\end{figure}

\section{Conclusions}
\label{sec:Conclusions}
In this paper, we have devised a low-complexity numerical algorithm to calculate the distribution of the path distance between a randomly selected road intersection and the $k$-th nearest node of a Cox point process driven by the Manhattan Poisson line process. This algorithm can be used to identify the minimum required density of facilities (modeled by a \ac{MPLCP}), e.g., charge stations for electric vehicles, to ensure that a vehicle at an intersection can reach the nearest available facility within a target time under a probability constraint. The distance distributions derived in this paper can also be used to calculate the distribution of network load within a cell of a V2X system. Finally, using real road network data, we illustrated the enhanced performance of the MPLP as compared to the state-of-art distance distribution model using the two-dimensional \ac{PPP}. It is straightforward to incorporate into our approach path distance distributions towards a specific direction, e.g., south, north-east, etc. In the future, it would be interesting to extend our analysis and data collection with larger maps and different urban areas, investigating the independence of the path distance distributions from the underlying road infrastructure, when the network of facilities is sparse.

\end{document}